\documentclass[12pt,fleqn]{amsart}

\usepackage{amsthm}
\usepackage{graphicx}
\usepackage{color}
\usepackage{amsmath}
\usepackage{amssymb}
\usepackage{amscd}
\usepackage{latexsym}
\usepackage{graphics}
\usepackage{url}
\usepackage{verbatim}

\theoremstyle{plain}
\newtheorem{theorem}{Theorem}[section]
\newtheorem{lemma}[theorem]{Lemma}

\theoremstyle{definition}
\newtheorem{definition}[theorem]{Definition}

\theoremstyle{remark}

\def\I#1{{[\![#1]\!]}}            

\begin{document}

\title{An explicit universal cycle for the \\ $(n-1)$-permutations
  of an $n$-set}

\author[F. Ruskey]{Frank Ruskey}
\address{Dept. of Computer Science, University of Victoria, CANADA}
\thanks{Research supported in part by NSERC}
\urladdr{http://www.cs.uvic.ca/~ruskey}

\author[A. Williams]{Aaron Williams}
\address{Dept. of Computer Science, University of Victoria, CANADA}

\begin{abstract}
We show how to construct an \emph{explicit} Hamilton cycle in the
  directed Cayley graph $\overrightarrow{\mathrm{Cay}}( \{ \sigma_n, \sigma_{n-1} \} :
  \mathbb{S}_n )$, where $\sigma_k = (1\ 2\ \cdots \ k)$.
The existence of such cycles was shown by
  Jackson (Discrete Mathematics, \textbf{149} (1996) 123--129) but the
  proof only shows that a certain directed graph is Eulerian, and
  Knuth (Volume 4 Fascicle 2, Generating All Tuples and Permutations (2005))
  asks for an explicit construction.
We show that a simple recursion describes our Hamilton cycle and
  that the cycle can be generated by an iterative algorithm that uses
  $O(n)$ space.
Moreover, the algorithm produces each successive edge of the cycle in constant time;
  such algorithms are said to be \emph{loopless}.
\end{abstract}

\maketitle

\section{Introduction and motivation}

There are many proofs in the mathematical literature showing the existence of
  Hamilton cycles or Eulerian cycles in important families of graphs.
However, turning these proofs into efficient algorithms often represents a
  significant challenge.

An interesting case in point is the well-known De Bruijn cycle, which is a
  length $k^n$ circular string over a $k$-ary alphabet with the property
  that every length $n$ string occurs as a substring.
The existence of De Bruijn cycles is commonly presented in undergraduate
  discrete mathematics courses as a consequence of a certain graph
  being Eulerian.
However, it is not widely known how to efficiently generate a De Bruijn cycle.
In the authors' view two aspects of this question have particular importance.

\begin{itemize}
\item
\textbf{Space, not time, is the primary enemy.}
A na\"{\i}ve solution would be to build the
   graph and then use a Eulerian cycle algorithm to produce the cycle.
This will be practical for small values of $n$ and $k$, but for large
   values space will be the limiting factor long before time becomes a
   factor.
In general, we need to be able to generate the Hamilton or Eulerian cycle
  \emph{without} building the graph, or storing exponentially-long sublists.
There are algorithms for building De Bruijn cycles that use space $O(n)$.
The earliest of these is due to Fredricksen and Maiorana \cite{FredricksenMaiorana}
  and is presented in Knuth \cite{Knuth}.
\item
\textbf{The development of efficient algorithms reveals structure.}
It is often worthwhile to turn a proof into an algorithm,
  or to develop an alternate proof, because the process often results
  in a deeper structural understanding of the
  cycles being listed.
For example, the efficient algorithm due to Fredricksen and Maiorana is based
   on necklaces, Lyndon words, and is related to pattern-matching and Lyndon factorization.
\end{itemize}

As another example from the Hamiltom cycle domain,
  Eades, Hickey, and McKay \cite{EHR} considered the graph
  $G(n,k)$ whose vertices are all length $n$ bitstrings with
  density $k$ and where two bitstrings are joined by an edge if
  they differ by transposing two adjacent bits.
They showed that $G(n,k)$ is Hamiltonian if and only if
  $n$ is even and $k$ is odd.
The proof is inductive and relies on the fact that the graph
  has a spanning subgraph that is the prism of two ``combs."
However, it was not at all clear how to turn that proof into
  an efficient algorithm.
Eventually an algorithm that mimics the proof was found that
  uses $O(n)$ space and take time $O(1)$ per bitstring generated
  \cite{HoughRuskey}.

In the present paper we are considering the construction of a
  ``universal cycle" for the $(n-1)$-permutations of an $n$-set (which
  we take to be $\{1,2,\ldots,n\}$).
Here a \emph{universal cycle} is a circular string of length $n!$ what contains each of the
  $n!$ different $(n-1)$-permutations as a (contiguous) substring.
For example, 321312 is a such a universal cycle for $n = 3$,
  since its substrings are 32, 21, 13, 31, 12, and 23.

More general universal cycles were introduced by Chung, Diaconis, and Graham \cite{ChungDiaconisGraham}
  as a way of extending the de Bruijn cycle idea to combinatorial objects
  in general.
The existence of a universal cycle for the $k$-permutations of an
  $n$-set was shown by Jackson \cite{Jackson} when $k<n$.
His proof sets up a certain natural Eulerian graph, call it $J_{k,n}$, and shows that any
  Eulerian cycle in that graph corresponds to the required
  universal cycle.
However, no explicit construction of the cycle is indicated.
The problem for $k = n-1$ is discussed by Knuth \cite{Knuth} in Exercise 112
  of Section 7.2.1.2.
On page 121 of \cite{Knuth} we find the following quote:
\begin{quote}
``At least one of these cycles must almost surely be easy to describe and to compute,
  as we did for de Bruijn cycles in Section 7.2.1.1.  But no simple construction
  has yet been found.''
\end{quote}

The purpose of this paper is to provide such a description and computational method.
We will show how to construct a particular universal cycle.
Our algorithm takes space $O(n)$ and uses a constant
  amount of time between successive outputs of characters in the cycle.
To be precise regrading the space requirement: The algorithm uses a constant
  number of arrays, each with $O(n)$ indices, 
  and each storing integers of value at most $n$.
Similarly regarding time, we use a constant number of operations
  (comparisons, increments, decrements, and parity tests)
  on integers of value at most $n$.

Universal cycles for the permutations of an $n$-set are not directly possible unless $n \leq 2$.
However, every $(n-1)$-permutation of an $n$-set can be uniquely extended to a permutation of an $n$-set by
  appending the unique missing symbol.
Thus, universal cycles for $(n-1)$-permutations can be viewed as universal cycles for permutations.
For example, $321312$ produces the permutations $32\underline{1}$, $21\underline{3}$, $13\underline{2}$,
$31\underline{2}$, $12\underline{3}$, and $23\underline{1}$, where the appended missing symbols are underlined.
For this reason, our results add to the already sizeable
  literature on generating permutations.
A good survey is provided by Sedgewick \cite{Sedgewick} and more recent
  developments are to be found in Knuth \cite{Knuth}.

We don't expect our algorithms to be a fast way to generate permutations using
  the usual model of computation, since at least $n-1$ of the $n$ values
  change at each step.
However, they will be fast if a circular representation is used; for example,
  when using linked lists or a circular array.
In a circular array we maintain a start position and do arithmetic on indices
  mod $n$.
They will also be fast if the permutation is stored as a computer word.
For example, we can store the permutations up to $n = 16$ by dividing 64 bit
  words into 4 half-bytes.
The shifts can then be accommodated in a few machine instructions.

Finally, we mention that additional symbols can
also be used to create universal cycles whose substrings are \emph{order isomorphic} to
permutations. For example, $421423$ produces the permutations $321$, $213$, $132$, $312$, $123$,
and $231$. Recently Johnson \cite{Johnson} proved a conjecture in \cite{ChungDiaconisGraham}
   by showing that $n+1$ symbols are always sufficient for constructing these universal cycles.

The paper is organized as follows.
In Section 2 we give our explicit construction as a certain recursively defined string.
Then, in Section 3, we show that this string can be generated by an algorithm that uses only a
  constant amount of computation between the output of successive symbols of the string ---
  the first such algorithm for a universal cycle.
In Section 4, we give further properties of our recursive construction;
  first some results on the number of $\sigma_n$ or $\sigma_{n-1}$ operations that are used,
  then that our ordering has an efficiently computable ranking function, and finally that
  it is ``multiversal," in a sense to be described later.
We conclude with Section 5, which contains some open problems.

\section{An explicit construction}

Initially, we will couch our discussion in terms of finding Hamilton paths
  in certain directed Cayley graphs.
Cayley graphs are denoted $X = \overrightarrow{\mathrm{Cay}}( \{ \alpha_1, \alpha_2, \cdots,  \alpha_k \} : \mathbb{G} )$.
Here $\{ \alpha_1, \alpha_2, \cdots,  \alpha_k \}$ is a generating set of a group $\mathbb{G}$.
The vertices of $X$ are the elements of $\mathbb{G}$ and the edges are all of the form
  $g \rightarrow \alpha_i g$; these edges are usually thought of as being labelled
  with $\alpha_i$.
In an \emph{undirected} Cayley graph, if $\alpha$ is in the generating set, then its
  inverse $\alpha^-$ is also in the generating set.
Driven by the question of Lov\'{a}sz of whether there is a Hamilton cycle in all \emph{undirected}
  Cayley graphs, there is a significant literature of results about Hamilton cycles in
  Cayley graphs.
A survey may be found in Gallian and Witte \cite{GallianWitte};
  see also Pak and Rado\v{s} Radoi\v{c}i\'{c} \cite{PakRado}.

In the solution to Exercise 112 of Section 7.2.1.2 Don Knuth implicitly
  poses the problem of finding an explicit expression for universal cycles of $(n-1)$-permutations
  of an $n$-set \cite{Knuth}.
This problem
  is equivalent to generating permutations of an $n$-set by
  rotations of the form $(1\ 2\ \cdots \ n)$ or $(1\ 2\ \cdots \ n{-}1)$;
  i.e., it is equivalent to asking whether the Cayley graph
\[
\Xi_n := \overrightarrow{\mathrm{Cay}}( \{ \sigma_n, \sigma_{n-1} \} : \mathbb{S}_n )
\]
  is Hamiltonian.
We use $\mathbb{S}_{k,n}$ to denote the set of $k$-permutations of the $n$-set $[n] = \{1,2,\ldots,n\}$.
In the case where $k = n$ we use $\mathbb{S}_n$.
Although we do not use this fact below, it is interesting to note that a short proof reveals that
  the graph $\Xi_n$ is the \emph{line graph} of the Jackson graph $J_{n-1,n}$.

Consider the binary string $S_n$ defined by the following recursive rules.
The base case is $S_2 = 00$.
Let $S_n = x_1 x_2 \cdots x_{n!}$ where $\overline{x}$ denotes flipping the bit $x$.
Then, for $n > 2$,
\begin{equation}
S_{n+1} := 001^{n-2}\ \overline{x}_1 001^{n-2}\ \overline{x}_2 \cdots 001^{n-2}\ \overline{x}_{n!}.
\label{eq:recur}
\end{equation}
We use above the usual convention that if $w$ is a string and $m$ is an integer then
  $w^m$ is $w$ concatenated together $m$ times, $w^m = ww \cdots w$; also
  $w^0$ is the empty string.

Below we list $S_3$, $S_4$, and $S_5$.
Each $S_i$ is of the form $ww$ since 00 has this property and the recurrence
  (\ref{eq:recur}) preserves it.
\begin{align*}
S_3 = \ \ \ & 00\ \bar{0}\ 00\ \bar{0}\ \ \   = \ \ \ 00\ 1\ 00\ 1. \\
S_4 = \ \ \ & 001\ \bar{0}\ 001\ \bar{0}\ 001\ \bar{1}\ 001\ \bar{0}\ 001\ \bar{0}\ 001\ \bar{1} = 001\ 1\ 001\ 1\ 001\ 0\ 001\ 1\ 001\ 1\ 001\ 0. \\
S_5 = (\ & 0011\ 1\ 0011\ 1\ 0011\ 0\ 0011\ 0\ 0011\ 1\ 0011\ 1\ \\
          & 0011\ 0\ 0011\ 0\ 0011\ 1\ 0011\ 1\ 0011\ 0\ 0011\ 1\ )^2 \\
\end{align*}

Now define the mapping $\phi$ by $0 \rightarrow \sigma_n$ and $1 \rightarrow \sigma_{n-1}$
  where $\sigma_k = (1\ 2\ \cdots \ k)$.

\begin{theorem}
The list $\phi(S_n)$ is a Hamilton cycle in
  the directed Cayley graph $\Xi_n$.
\end{theorem}

\begin{proof}
In listing the Hamilton cycle we use one-line notation for the permutations, starting
  with $n\ n{-}1\ \cdots\ 2\ 1$, and think of the cycles $\sigma_{n-1}$ and $\sigma_{n}$
  as acting on the positions in the one-line notation.
Thus, in a slight abuse of notation,
\[
\phi(S_3) = 321, 213, 132, 312, 123, 231,
\]
since $S_3$ implies the successive application of
  $\sigma_3$, $\sigma_3$, $\sigma_2$, $\sigma_3$, $\sigma_3$, and finally $\sigma_2$ to
  map the last permutation to the first.

Our proof strategy is to give an explicit listing of permutations of $[n]$ with the
  required properties and then show that it is equivalent to (\ref{eq:recur}).
Recursively define a \emph{circular} list $\Pi(n) = \Pi(n)_0,\Pi(n)_1,\ldots,\Pi(n)_{n!-1}$
  of permutations of $[n]$.
For small values of $n$, define $\Pi(1) = 1$, $\Pi(2) = 21,12$, and $\Pi(3) = \phi(S_3)$.
Every $n$-th permutation of $\Pi(n)$ is defined as follows.
\begin{equation}
\Pi(n)_{jn} := n \Pi(n-1)_j.
\label{eq:n-th}
\end{equation}
The $n-1$ permutations that follow $n \pi$, where $\pi = \Pi(n-1)_j$, are defined to be
\begin{equation}
\sigma_n (n\pi),\ \sigma_n^2 (n\pi),\
\sigma_{n-1} (\sigma_n^2  (n\pi)),\ \ldots,\ \sigma_{n-1}^{n-3} (\sigma_n^2 (n\pi)).
\label{eq:nextones}
\end{equation}
The list $\Pi(4)$ is shown in Table \ref{table:count}, column (d).
The permutation $n \pi$ followed by the permutations above comprise the sublist
  $\Pi(n)_{jn}, \Pi(n)_{jn+1}, \ldots, \Pi(n)_{(j+1)n - 1}$
  and these permutations are all distinct since the position of $n$ is
  successively in the $n$ different positions $1$, $n$, $n-1$, \ldots, $2$.
Furthermore, because we can recover $\pi$ from any permutation in this sublist,
  the uniqueness of every permutation in $\Pi(n)$ follows
  inductively from the uniqueness of every permutation in $\Pi(n-1)$.

It remains only to prove that successive permutations differ by $\sigma_n$ or
  $\sigma_{n-1}$ and that the list is circular.
It is clear from (\ref{eq:nextones}) that successive permutations differ by
  $\sigma_n$ or $\sigma_{n-1}$, except
  for those that precede the one of the form $n\pi$ successively followed by $n\pi$.
Let $a \tau z$ be a permutation of $1,2,\ldots,n-1$ where $a$ and $z$ are numbers and
  $\tau$ is a sequence (of length $n-3$).
Note that the last permutation of (\ref{eq:nextones}) is
\[
\sigma_{n-1}^{n-3} ( \sigma_n^2 ( na \tau z )) = \sigma_{n-1}^{-2} ( \sigma_n^2 ( na \tau z ))
  = \sigma_{n-1}^{-2} ( \tau z na ) = zn\tau a.
\]
Now suppose that $\pi = \Pi(n-1)_j = a \tau z$ and $\pi' = \Pi(n-1)_{j+1}$.
Inductively, either $\pi' = \sigma_{n-1}(\pi)$ or $\pi' = \sigma_{n-2}(\pi)$.
Observe that
\begin{gather}
\label{eq:10}
\sigma_n    ( zn\tau a) = n\tau az = n \sigma_{n-2}( a\tau z ), \text{ and } \\
\label{eq:01}
\sigma_{n-1}( zn\tau a) = n\tau za = n \sigma_{n-1}( a\tau z ).
\end{gather}
Since the successor of $a\tau z$ is either $\sigma_{n-2}( a\tau z )$ or $\sigma_{n-1}( a\tau z )$,
  the transition to the permutation $\pi'$ is also of the correct form; successive permutations
  in $\Pi(n)$ differ by $\sigma_n$ or $\sigma_{n-1}$.
The circularity of the list follows inductively from the circularity of the list $\Pi(n-1)$
  (alternatively we could use Lemma \ref{lemma:circular} below).
Furthermore, in terms of the mapping $\phi$ defined earlier, the bits
  are flipped; a 0 ($\sigma_{n-1}$) transition in $\Pi(n-1)$ becomes a
  1 ($\sigma_{n-1}$) transition in $\Pi(n)$ by (\ref{eq:01}), and a
  1 ($\sigma_{n-2}$) transition in $\Pi(n-1)$ becomes a
  0 ($\sigma_{n}$) transition in $\Pi(n)$ by (\ref{eq:10}).
\end{proof}

\begin{lemma}
Any Hamilton path
  in $\Xi_n$ 
  is, in fact, a Hamilton cycle.
\label{lemma:circular}
\end{lemma}

\begin{proof}
Suppose that $\Pi = \Pi_1,\Pi_2,\ldots,\Pi_{n!}$ is a Hamilton path in
  $\Xi_n$ 
  that is not a Hamilton cycle.
In particular $\sigma_n(\Pi_{n!}) \neq \Pi_1$ and
  $\sigma_{n-1}(\Pi_{n!}) \neq \Pi_1$.
Thus $\Pi_{n!} \neq \sigma_n^-( \Pi_1 )$ and $\Pi_{n!} \neq \sigma_{n-1}^-( \Pi_1 )$.
We must then have that $\sigma_n^-( \Pi_1 ) \rightarrow \sigma_{n-1} (\sigma_n^-( \Pi_1 ))$ and
  $\sigma_{n-1}^-( \Pi_1 ) \rightarrow \sigma_n (\sigma_{n-1}^-( \Pi_1 ))$ are distinct edges in
  $\Pi(n)$.
However, an easy calculation shows that
  $\sigma_{n-1}^- \sigma_n = \sigma_n^- \sigma_{n-1} = (n{-}1\ n)$
  and thus these permutations are identical.
This contradiction shows that $\Pi$ is a Hamilton cycle.
\end{proof}

The proof shows that in fact the lemma is true for any Cayley graph on
  two generators $\rho$ and $\tau$ for which $\tau^- \rho$ is an involution.

The universal cycle for $(n-1)$-permutations of $[n]$ is obtained by
  recording the first symbol in each of the permutations in $\Pi(n)$.
We use $U_n$ to denote the resulting universal cycle.

\section{A loopfree algorithm}

Suppose that in our recurrence (\ref{eq:recur}) for $S_{n+1}$ that for each ``new" bit we record
  the value $n$, and apply this idea recursively.
Call the corresponding new sequence $R_{n+1}$.
That is, $R_2 = 11$, and for $n > 1$,
\[
R_{n+1} = n^n y_1 n^n y_2 \cdots n^n y_{n!},
\]
where $R_n = y_1 y_2 \cdots y_{n!}$.
For example
\[
R_4 = 333\ 2\ 333\ 2\ 333\ 1\ 333\ 2\ 333\ 2\ 333\ 1.
\]
The sequence $R_4$ is exactly the sequence that is obtained by recording the most significant
  position that changes when counting with the multi-radix numbers
  with parameters $2 \times 3 \times 4$, when the numbers are indexed 1, 2, 3, from
  left-to-right.
See Table \ref{table:count}, columns (a) and (b).
In general, $R_{n}$ gives us the positions when counting with
  multi-radix numbers $2 \times 3 \times \cdots \times n$.

\begin{table}
\begin{tabular}{ccccccr}
(a) & (b) & (c) & (d) & (e) & (f) & (g) \\
234 & $R_4$ & 234 & $S_4$ &  $\mathbb{S}_4$ & $U_4$ & rank \\ \hline
000 & 3 & 000 & . . 0 & 4321 &  4 & 0 \\
001 & 3 & 001 & . . 0 & 3214 &  3 & 1 \\
002 & 3 & 002 & . . 1 & 2143 &  2 & 2 \\
003 & 2 & 003 & . 1 . & 1423 &  1 & 3 \\
010 & 3 & 013 & . . 0 & 4213 &  4 & 4 \\
011 & 3 & 012 & . . 0 & 2134 &  2 & 5 \\
012 & 3 & 011 & . . 1 & 1342 &  1 & 6 \\
013 & 2 & 010 & . 1 . & 3412 &  3 & 7 \\
020 & 3 & 020 & . . 0 & 4132 &  4 & 8 \\
021 & 3 & 021 & . . 0 & 1324 &  1 & 9 \\
022 & 3 & 022 & . . 1 & 3241 & 3 & 10 \\
023 & 1 & 023 & 0 . . & 2431 & 2 & 11 \\
100 & 3 & 123 & . . 0 & 4312 & 4 & 12 \\
101 & 3 & 122 & . . 0 & 3124 & 3 & 13 \\
102 & 3 & 121 & . . 1 & 1243 & 1 & 14 \\
103 & 2 & 120 & . 1 . & 2413 & 2 & 15 \\
110 & 3 & 110 & . . 0 & 4123 & 4 & 16 \\
111 & 3 & 111 & . . 0 & 1234 & 1 & 17 \\
112 & 3 & 112 & . . 1 & 2341 & 2 & 18 \\
113 & 2 & 113 & . 1 . & 3421 & 3 & 19 \\
120 & 3 & 103 & . . 0 & 4231 & 4 & 20 \\
121 & 3 & 102 & . . 0 & 2314 & 2 & 21 \\
122 & 3 & 101 & . . 1 & 3142 & 3 & 22 \\
123 & 1 & 100 & 0 . . & 1432 & 1 & 23 \\
\end{tabular}
\caption{(a) Counting in multi-radix base $2 \times 3 \times 4$, (b) the $R_4$ sequence,
  (c) the corresponding multi-radix Gray code, (d) indented version of $S_4$, (e) the list $\Pi(4)$, (f) the
  universal cycle $U_4$, and (g) the rank of each permutation.}
\label{table:count}
\end{table}

These observations suggest that we may be able to efficiently generate
  the $S_{n}$ sequence by modifying the
  classic algorithm for counting with multi-radix numbers.
In the classic algorithm the multi-radix number is stored in the array
  $a_{n-1} \cdots a_2 a_1$ and $j$ is used to represent the rightmost, or smallest,
  index where $a_j$ is not at its maximum value.
The next multi-radix number is obtained by incrementing $a_j$ and setting all
  values to its right to 0.
Now suppose that we just incremented the integer in position $j$ so that the multi-radix number is
  $a_{n-1} \cdots a_j a_{j-1} \cdots a_1 = a_{n-1} \cdots a_j 0 \cdots 0$.
Then the corresponding $R_n$ value is $n-j$ and so the non-recursive part of
  the $S_j$ sequence that we are listing is going through the pattern
  $001^{n-j-1}$ or the pattern $110^{n-j-1}$, depending on whether $j$ is
  odd or even, respectively.
For proposition $P$ we use the notation $\I{P}$ to mean the value $1$ if
  $P$ is true and the value $0$ if $P$ is false; also
  $\oplus$ denotes exclusive-or.
The expression $\I{ j \text{ even } \oplus a_j \le 1 }$ gives the correct
  value of the bit to be output.
Below is the entire algorithm, rendered in pseudo-code.

\begin{center}
\fbox{\parbox[b]{.99\linewidth}{
\begin{tabbing}
XXXX \= XXXX \= XXXX \kill
\> $a_{n{+}1} a_n \cdots a_1 \leftarrow 0 \ 0 \ \cdots \ 0$; \\
\> \textbf{repeat} \\
\> \> $j \leftarrow 1$; \\
\> \> \textbf{while} $a_j = n-j$ \textbf{do}\  $a_j \leftarrow 0$;\ $j \leftarrow j+1$;\ \textbf{od}; \\
\> \> \textsf{output}( $\I{ j \text{ even } \oplus a_j \le 1 }$ ); \\
\> \> $a_j \leftarrow a_j + 1$; \\
\> \textbf{until} $j \ge n$;
\end{tabbing}
}}
\end{center}

There is an loopless algorithm for listing multi-radix numbers as a Gray code
  in which the value in only one position changes and that change is by $\pm 1$
  (see, for example, Williamson \cite{Williamson}, pg.~112, or Knuth \cite{Knuth}, pg.~20).
Together with the ideas used in the previous ``counting" algorithm,
  we can adapt those loopless algorithms to get a loopless algorithm for generating
  $S_n$ or our universal cycle.
In the Gray code for multi-radix numbers, the values in a given position
  alternately increase and decrease.
Furthermore, the values change in exactly the positions given by the $R_n$ sequence.
In the implementation we maintain a direction array $d$ where $+1$ means increase
  and $-1$ means decrease
We also maintain an array $f$ of ``focus pointers" which
  allow instant access to the next position whose value will change (we set $f_n = n{+}1$
  (instead of $n$) so that the last iteration is handled correctly).
See Table~\ref{table:count}, column (c), for an example.

Thus the values of $j$ from the counting algorithm are exactly the same
  in the Gray code algorithm, except that in the Gray code algorithm $j$ is
  the position where a value changes.
The only complication arises because the values in a given position can
  be decreasing, and so the test ``$a_j \le 1$" is not sufficient.
Fortunately, all algorithms that looplessly implement the Gray code
  maintain an array of directions $d_{n-1} \cdots d_2 d_1$ for each
  position, where $d_i \in \{+1,-1\}$, indicating whether the values in
  that position are currently increasing $(+1)$ or decreasing $(-1)$.
If $d_j = +1$ then we can continue to test $a_j \le 1$, but to
  account for $d_j = -1$, we need to test
\[
(a_j \le 1 \ \textbf{ and } \ d_j = 1) \ \textbf{ or } \ (a_j \ge n-j-1 \ \textbf{ and } \ d_j = -1).
\]

We can ``optimize" this condition.  Notice that the test
  ($a_j \le 1 \ \textbf{ and } \ d_j = 1$) can be replaced by $a_j - d_j \le 0$.
This change is possible because if $d_j = -1$ then $a_j - d_j$
  is guaranteed to be greater than zero because $a_j \geq 0$.
Therefore, if $a_j - d_j \le 0$, then this immediately implies that $d_j=1$ and so $a_j - 1 \le 0$,
  which is equivalent to the original test $a_j \le 1$.
Likewise, the test ($a_j \ge n-j-1 \ \textbf{ and } \ d_j = -1$) can be replaced by $a_j - d_j \geq n-j$.
Below is our loopless algorithm in full detail.

\begin{center}
\fbox{\parbox[b]{.99\linewidth}{
\begin{tabbing}
XXXX \= XXX \= XXXX \kill
\> $a_{n+1} a_n \cdots a_1 \leftarrow   0 \ 0\ 0\ \cdots \ 0$; \\
\> $d_{n} d_{n-1 }\cdots d_1 \leftarrow 1 \ 1\ 1\ \cdots \ 1$; \\
\> $f_{n} f_{n-1} \cdots f_1 \leftarrow n{+}1\ n{-}1\ n{-}2\ \cdots \ 1$; \\
\> \textbf{repeat} \\ 
\> \> $j \leftarrow f_1$;\ $f_1 \leftarrow 1$; \\
\> \> \textsf{output}( \I{ $j$ even $\oplus$ $(a_j - d_j \le 0 \ \textbf{ or } \ a_j - d_j \ge n-j)$ } ); \\
\> \> $a_j \leftarrow a_j + d_j$; \\
\> \> \textbf{if} $a_j = 0$ \textbf{or} $a_j = n-j$ \textbf{then}\  $d_j \leftarrow -d_j$;\ $f_j \leftarrow f_{j+1}$;\  $f_{j+1} \leftarrow j+1$;\ \textbf{fi}; \\
\> \textbf{until} $j \ge n$;
\end{tabbing}
}}
\end{center}

It is also possible to output the universal cycle itself in a loopless
  manner, but an additional circular array is required to hold the
  current permutation.
To follow are the details.
Define an array $\pi_1 \pi_2 \cdots \pi_n$ initialized to
  $n\ n{-}1\ \cdots \ 1$ and an index $t$ that will be incremented
  mod $n$ on each iteration of the algorithm.
We will think of $\pi$ as a circular array.
The index $t$ is the position of the last element of $\pi$,
  so initially $t = n$.
As each bit of $S_n$ is determined, we will ouput the
  first element of $\pi$ (i.e., the one in position $t+1$).
If the bit is a 1, so that $\sigma_{n-1}$ is acting on $\pi$
  then we need to swap the last two elements:
  $\pi_{t-1} \leftrightarrow \pi_t$. 
In other words the \textsf{output} statements in the preceding code fragments is
  replaced with the following code where $\mathit{expr}$ is the
  expression inside of the \textsf{output} statement in either the previous counting
  algorithm of the previous loopless algorithm.

\begin{center}
\fbox{\parbox[b]{.99\linewidth}{
\begin{tabbing}
XXXX \= XXXX \= XXXX \kill
\> $t' \leftarrow t$; \ \  $t \leftarrow (t+1) \bmod{n}$; \\
\> \textsf{output}( $\pi_{(t+1) \bmod{n}}$ ); \\
\> \textbf{if} $\mathit{expr} = 1$ \textbf{then} $\pi_t \leftrightarrow \pi_{t'}$ \textbf{fi};
\end{tabbing}
}}
\end{center}

Finally, we note that every permutation can be output in a circular fashion by
  outputting $\pi$ and $t$.
We could also use a linked list, which would give a loopless permutation
  generation algorithm.

\section{Further properties}

In this section we explore further properties of $\Xi_n$ and our Hamilton cycle.

\subsection{How many of each rotation is used?}

It is clear from the recurrence relation (\ref{eq:recur}) that the
  number, call it $f_n$, of $\sigma_{n}$'s in $\phi(S_n)$ satisfies
  the recurrence relation
\begin{equation}
f_{n+1} = \begin{cases}
2 & \text{ if } n = 1 \\
3n!-f_n & \text{ if } n > 1.
\end{cases}
\end{equation}
This recurrence relation can be iterated to obtain
\[
f_n = 2(-1)^n - 3 \sum_{k=1}^{n-1} (-1)^k(n-k)!,
\]
from which it follows that
\[
f_n \sim 3(n-1)! \ \ \ \text{ or }\ \ \  \frac{f_n}{n!} \sim \frac{3}{n}.
\]
Interestingly, this sequence appears in OEIS \cite{OEIS} as A122972$(n+1)$ as the
  solution to the ``symmetric" recurrence relation $a(n+1) = (n-1) \cdot a(n) + n \cdot a(n-1)$.
The values of $f_n$ for $n = 1..10$ are 1, 2, 4, 14, 58, 302, 1858, 13262, 107698, 980942.

Consider the cosets induced by $\sigma_{n}$; there are $n!/n = (n-1)!$ of them.
In a Hamilton cycle there must be at least one $\sigma_{n-1}$ edge that leaves each coset, and thus there
  must be at least $(n-1)!$ of them.
Alternatively, consider the cosets induced by $\sigma_{n-1}$; there are $n!/(n-1) = n \cdot (n-2)!$ of them.
In a Hamilton cycle there must be at least one $\sigma_n$ edge that leaves the coset, and thus there
  must be at least $n \cdot (n-2)!$ of them.  We can make a stronger statement regarding the
  $\sigma_{n}$ edges.

\begin{lemma}
The least number of $\sigma_n$ edges in any Hamilton cycle in
  $\Xi_n$ is $2n(n-2)!-2$.
\label{lemma:least}
\end{lemma}
\begin{proof}
First, observe that
\[
\sigma_{-1}^- \sigma_{n} \sigma_{n-1}^- \sigma_{n} = (n{-}1\ n)(n{-}1\ n) = \mathit{id}.
\]
The two $\sigma_n$ edges above are incident with the same unordered pair of cosets induced
  by $\sigma_{n-1}$.
Thus if we contract each coset into a singe super-vertex, then the resulting graph, call
  it $Q_n$, is undirected in the sense that every directed edge is paired with an edge
  in a 2-cycle.
Furthermore, it is not hard to see that if one of those $\sigma_n$ edges is used in
  a Hamilton cycle, then so must the other.
Thus a Hamilton cycle in $X_n$ becomes a connected spanning subgraph of $Q_n$.
Since a minimal connected spanning subgraph is a spanning tree, and any spanning
  tree has $n \cdot (n-2)! -1 $ edges, the number of $\sigma_n$ edges is at
  least $2n(n-2)!-2$.
\end{proof}

\begin{figure}
\includegraphics[width=4.5in]{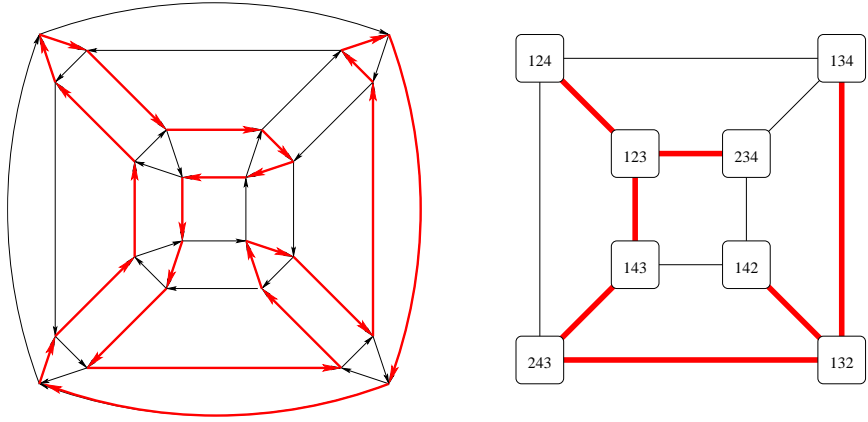}
\caption{The Cayley graph $\Xi_4$ on the left.  The graph $Q_4$ on the right.
  The thick (red) edges indicate the Hamilton cycle $S_4$.}
\label{fig:Cayley}
\end{figure}

Figure \ref{fig:Cayley} shows the Cayley graph $X_4$.
Note that the contracted graph $Q_4$ is the 3-cube.
The red edges show the Hamilton cycle $S_4$.
In this case $S_n$ corresponds to a spanning tree in $Q_n$,
  but this is not the case for $n \ge 6$.

\subsection{Ranking}

The \emph{rank} of a permutation $\pi$ is the value $r$ for which $\Pi(n)_r = \pi$.
Our recursive equation for the rank depends on the position of $n$
  within the permutation being ranked.
  From the definition of $\Pi(n)$ we can infer that

\begin{gather*}
\mathrm{rank}( a_1 a_2 \cdots a_{k-1} n a_{k+1} \cdots a_n ) \\ =
\begin{cases}
   0 & \text{ if } n = 1, \\
   n \cdot \mathrm{rank}( a_2 a_3 \cdots a_{n} ) & \text{ if } k = 1, \\
   n-k+1 + n \cdot \mathrm{rank}( a_n a_{k+1} \cdots a_{n-1} a_1 \cdots a_k ) & \text{ if } k > 1.
\end{cases}
\end{gather*}

The expression $n-k+1$ accounts for the position of the $n$, and the rest comes
  from the recursive part of the definition of $\Pi(n)$.
We can also express the rank as

\begin{equation*}
\mathrm{rank}(\ \alpha n \beta\ ) =
\begin{cases}
   0 & \text{ if } \alpha = \beta = \epsilon, \\
   n \cdot \mathrm{rank}(\ \beta\ ) & \text{ if } \alpha = \epsilon, \\
   n-|\alpha| + n \cdot \mathrm{rank}(\ \sigma( \beta ) \alpha\ ) & \text{ otherwise },
\end{cases}
\end{equation*}
where $\sigma(\beta)$ is $\beta$ rotated one position to the right.

Implemented in the obvious manner, these recurrence relations lead to algorithms
  that use $O(n^2)$ arithmetic operations on integers as large as $n!$.

\subsection{Multiversal Cycle Property}

In this section we prove that the sequence $\Pi(n) = \Pi(n)_0, \Pi(n)_1, \ldots, \Pi(n)_{n!}$,
  written out as a long string of symbols by concatenating each permutation,
  is a ``multiversal cycle".
We denote this ``flattening" of $\Pi(n)$ as $\coprod(n)$.
For example, consider $\coprod(3) = 321\ 213\ 132\ 312\ 123\ 231$.
Starting in positions 0,1, or 2 and advancing the position in increments of 3,
  recording the first two symbols, we obtain

\begin{center}
\begin{tabular}{c|cccccc}
0 & 32 & 21 & 13 & 31 & 12 & 23 \\ \hline
1 & 21 & 13 & 32 & 12 & 23 & 31 \\ \hline
2 & 12 & 31 & 23 & 21 & 32 & 13
\end{tabular}
\end{center}

In each case a complete set of all 2-permutations of [3] is obtained.
The purpose of this section is to prove that this property holds in general.

\begin{definition}
A \emph{multiversal} cycle for the $(n-1)$-permutations of an $n$-set is
a circular string $a_0 a_1 \cdots a_{N-1}$ of length $N = n \cdot n!$ such
that, for all $m = 0,1,\ldots,n-1$,
\begin{equation}
\{ a_{m+in} \cdots a_{m+in+n-2} \mid i = 1,2,\ldots,n! \} = \mathbb{S}_{n{-}1,n},
\label{eq:multidef}
\end{equation}
where arithmetic in the indices is taken mod $n$.
\end{definition}

Before getting to the main theorem in this section we prove a
  technical lemma.
\begin{lemma}
For all $i \neq 0,-1 \bmod{n}$, if $\coprod(n) = a_0 a_1 \cdots a_{N-1}$, then
\[
a_{i} = a_{i+n-1}.
\]
\label{lemma:tech}
\end{lemma}
\begin{proof}
Because $i \neq 0 \bmod{n}$ the numbers $a_{i}$ and $a_{i+n-1}$
  lie in two successive permutations of $\Pi(n)$.
The conclusion now follows since successive
  permutations differ by $\sigma_n$ or $\sigma_{n-1}$.
The $i \neq -1 \bmod{n}$ condition is necessary when they differ by $\sigma_{n-1}$.
\end{proof}

\begin{theorem}
The string $\coprod(n) = a_0 a_1 \cdots a_{n!-1}$ is a multiversal cycle.
\label{thm:multiversal}
\end{theorem}
\begin{proof}
The proof is by induction on the value $m$ in the definition.
The base case $m = 0$ satisfies (\ref{eq:multidef}) because
  $\Pi(n)_0, \Pi(n)_1, \ldots, \Pi(n)_{n!-1}$ is a listing of all
  permutations of $[n]$, so ignoring the last character of each
  permutation gives a complete listing of all $(n-1)$-permutations
  of $[n]$.
Similarly, when $m = 1$, ignoring the first character of each
  permutation also gives a complete listing of all $(n-1)$-permutations
  of $[n]$.
We now argue by contradiction.
Suppose that there are some values $m > 1$, $i$ and $i'$, with $i \neq i'$, such that
\begin{equation}
a_{m+in} \cdots a_{m+in+n-2} = a_{m+i'n} \cdots a_{m+i'n+n-2}.
\label{eq:multiproof}
\end{equation}
Inductively, we know that
\[
a_{m-1+in} \cdots a_{m-1+in+n-2} \neq a_{m-1+i'n} \cdots a_{m-1+i'n+n-2}.
\]
Thus it must be the case that $a_{m-1+in} \neq a_{m-1+i'n}$.
However, applying Lemma \ref{lemma:tech} to $a_{m-1+in}$ and $a_{m-1+i'n}$ gives
\[
a_{m-1+in} = a_{m-1+in+n-1} \text{ and } a_{m-1+i'n} = a_{m-1+i'n+n-1},
\]
so long as $m \neq 0,1$.
But by (\ref{eq:multiproof}) we now have
\[
a_{m-1+in} = a_{m-1+in+n-1} = a_{m+in+n-2} = a_{m+i'n+n-2} = a_{m-1+i'n+n-1} = a_{m-1+i'n},
\]
which is a contradiction.
\end{proof}

The careful reader will have noted that Lemma \ref{lemma:tech} and Theorem \ref{thm:multiversal}
  apply to \emph{any} Hamilton cycle in $\Xi_n$
  since the only property that we use is that successive permutations
  differ by $\sigma_{n-1}$ or $\sigma_n$.

\section{Final Remarks, Open Problems}

In this paper we have developed an explicit algorithm for generating a universal cycle
  for the $(n-1)$-permutations of an $n$-set.
This is the first universal cycle for which a loopless algorithm has been discovered.

Below is a list of open problems inspired by this work.
\begin{itemize}
\item
Can the results of this paper be extended to $k$-permutations of $[n]$
  for $1 \le k < n-1$?
\item
Among all Hamilton cycles in $\Xi_n$ we determined in Lemma \ref{lemma:least}
  the least number of $\sigma_n$ edges that need to be used in a
  Hamilton cycle in $\Xi_n$.
What is the least number of $\sigma_{n-1}$ edges that need be used?
In our construction, the number of $\sigma_n$ edges is asymptotic to $3/n$ and the
  number of $\sigma_{n-1}$ edges is asymptotic to $(n-3)/n$.
Is there a general construction that uses more $\sigma_n$ edges than
  $\sigma_{n-1}$ edges?
\item
Can the results of this paper be extended to the permutations of a multiset?
That is, given multiplicities $n_0, n_1, \ldots , n_t$, where $n_i$ is the
  number of times $i$ occurs in the multiset
  and $n = n_0 + n_1 + \cdots + n_t$, is there a circular string $a_1 a_2 \cdots a_N$
  of length $N = {N \choose n_0,n_1,\ldots,n_t}$ with the property that
\[
\{ a_i\ a_{i+1}\ \cdots\ a_{i+n-2}\ \iota(a_i,a_{i+1},\ldots,a_{i+n-2}) \mid 1 \le i \le N \}
\]
  is equal to the set of all permutations of the multiset.
Since the length of $a_i a_{i+1} \cdots a_{i+n-2}$ is $n-1$ it is not a permutation
  of the multiset; one character is missing.
The function $\iota$ gives the missing character.
We call these strings \emph{shorthand universal cycles}.
The current paper gave a shorthand cycle for permutations of $[n]$.
\item
It would be interesting to gain more insight in to the ranking process.
Is there a way to iterate the recursion so that it can be expressed as a sum?
\end{itemize}

\end{document}